\documentclass[preprint,NumberedRefs,nolinenumbers]{JASA}
\usepackage{comment}
\usepackage{soul}
\usepackage[bottom]{footmisc} 
\newcommand\mbf{\mathbf}

\begin{document}

\title[Differentiable physics for sound field reconstruction]{Differentiable physics for sound field reconstruction}

\author{Samuel A. Verburg}
\email{saveri@dtu.dk}
\affiliation{Department of Electrical and Photonics Engineering, Technical University of Denmark (DTU), \O rsteds Plads B343, 2800 Kgs. Lyngby, DK}

\author{Efren Fernandez-Grande}
\affiliation{Departamento de Ingeniería Audiovisual y Comunicaciones, Universidad Politécnica de Madrid (UPM), Madrid, Spain}

\author{Peter Gerstoft}
\affiliation{Department of Electrical and Photonics Engineering, Technical University of Denmark (DTU), \O rsteds Plads B343, 2800 Kgs. Lyngby, DK}

\begin{abstract}
Sound field reconstruction involves estimating sound fields from a limited number of spatially distributed observations. This work introduces a differentiable physics approach for sound field reconstruction, where the initial conditions of the wave equation are approximated with a neural network, and the differential operator is computed with a differentiable numerical solver. The use of a numerical solver enables a stable network training while enforcing the physics as a strong constraint, in contrast to conventional physics-informed neural networks, which include the physics as a constraint in the loss function. 
We introduce an additional sparsity-promoting constraint to achieve meaningful solutions even under severe undersampling conditions. Experiments demonstrate that the proposed approach can reconstruct sound fields under extreme data scarcity, achieving higher accuracy and better convergence compared to physics-informed neural networks.
\end{abstract}

\maketitle

\section{Introduction}
Sound field reconstruction refers to the inverse problem of estimating a sound field over time and space from a limited number of spatially distributed observations. Typically, the sound field is estimated using a suitable model to interpolate or extrapolate the measured data. A main challenge is the large number of measurements required, which increases with the domain size and the frequency. Therefore, efforts are devoted to developing efficient models to reconstruct sound fields with minimal data.

State-of-the-art sound field reconstruction relies on models that incorporate the physics of wave propagation.\cite{koyama2025physics,van2025deep} Linear models based on analytical solutions of the wave equation\cite{ueno2017sound,antonello2017room,ueno2018kernel,verburg2018reconstruction,koyama2019sparse,hahmann2022convolutional} approximate the sound field as the linear combination of physically-motivated functions. Therefore, the approximation abides by the physics of wave propagation by design. Sparsity-promoting representations are particularly relevant since they can greatly reduce the sampling requirements to estimate the sound field.\cite{mignot2013low,antonello2017room,verburg2018reconstruction,koyama2019sparse}
However, the choice of interpolation functions is critical, and linear models often lack representation power when the sound field does not conform with the chosen functions, making it difficult to find general sparsifying representations. Machine learning approaches, such as dictionary learning,\cite{hahmann2021spatial,damiano2024zero} deep learning,\cite{lluis2020sound,pezzoli2022deep,fernandez2023generative,bi2024point} kernel regression,\cite{caviedes2021gaussian,ribeiro2024sound,figueroa2025reconstruction} are often more flexible than linear combinations of elementary waves, but they are non-physical unless physical constraints are explicitly incorporated. In this context, physics-informed neural networks (PINNs) have emerged as a powerful framework to solve problems involving wave propagation.\cite{Moseley2020,Borrel2021,rasht2022physics,liu2024deep,liu2024spatial,yoon2024predicting,Olivieri2021,song2022versatile,karakonstantis2024room,ma2024sound} 
PINN integrate physical constraints in the form of partial differential equation (PDE) residuals, which are minimized along with a data-fitting term during training. PINNs achieve better generalization and data efficiency compared with purely data-driven models. On the downside, training PINNs is notoriously challenging, as the ill-conditioned differential operators in the PDE residual leads to slow convergence and failure to learn meaningful solutions.\cite{krishnapriyan2021characterizing,rathore2024challenges,wang2021understanding,wang2022and,basir2022critical} 

This study introduces a differentiable physics (DP) approach for sound field reconstruction. DP describes the integration of numerical PDE solvers into the training process of machine learning models.\cite{thuerey2021physics} The term \textit{differentiable physics} originates from the differentiable programming paradigm in which the variables of a computer program are made differentiable with respect of each other via automatic differentiation (AD).
In DP the unknown parameters of the physical system are approximated with a learnable model (e.g., a neural network) while the differential operators are computed using existing numerical methods, such as finite differences or finite elements.\cite{mitusch2021hybrid,zhu2021general} Formulating the PDE solver in a differentiable way makes training via gradient descent possible. 

At the core of DP is the use of AD to compute gradients through the solver. AD is a computational technique that enables the efficient evaluation of function derivatives by breaking it down into a sequence of elementary operations and applying the chain rule.\cite{baydin2018automatic} A computational graph keeps track of the operations and stores intermediate results, computing the function derivatives through a forward/backward pass. Both neural networks and PDE solvers can be viewed as a sequence of arithmetic operators for which AD compute gradients. DP has been applied to various domains like molecular dynamics, \cite{schoenholz2020jax} control of physical systems, \cite{de2018end,holl2020learning} computational fluid dynamics\cite{um2020solver}, robotics,\cite{heiden2021neuralsim} solid-solid interaction,\cite{newbury2024review} radio propagation,\cite{hoydis2024learning} etc. Software packages with DP capabilities include DiffTaichi,\cite{hu2019difftaichi} PhiFlow,\cite{holl2020phiflow} JAX MD,\cite{schoenholz2020jax}, Warp,\cite{warp2022} and j-Wave.\cite{stanziola2023j} In acoustic inverse problems, the computation of gradients through numerical solvers have been used to estimate of boundary impedance, \cite{antonello2015evaluation} wavefield inversion,\cite{zhu2021general} and source directivity. \cite{takeuchi2019source}

In our DP approach to sound field reconstruction, the initial condition is modeled with a neural network, and a differentiable finite difference solver is then used to solve the wave equation. We show that even if the network is trained for a given discretization, the sound field can be reconstructed at higher resolutions since the network can be queried at any point in the domain.
Furthermore, we propose a sparsity-promoting constraint to the initial condition that enables estimating sound fields with little data. In a series of experiments we show that the proposed DP approach is more robust and presents better convergence than PINNs, while achieving errors an order of magnitude smaller. The combination of DP and sparsity-promoting regularization provides accurate estimations even in challenging highly undersampled scenarios.

\section{Sound field reconstruction}
Let us consider the acoustic pressure field $p(\mathbf{r},t)$ in the spatio-temporal domain $\Omega \times [0, T]$, where $\Omega \subset \mathbb{R}^d$, $d \in 2, 3$, $\mathbf{r} \in \Omega$, and $t \in [0, T]$. The pressure field is the solution of the wave equation 
\begin{equation}
    \mathcal{D}[p] := \nabla^2 p(\mathbf{r},t) - \frac{1}{c^2}\frac{\partial^2 p(r,t)}{\partial t^2} = 0,
    \label{eq:wave_eq}
\end{equation}
with initial conditions, 
\begin{align}
    p(\mathbf{r},0) &= g(\mathbf{r}), \label{eq:initial_condition}\\
     \frac{\partial p}{\partial t}(\mathbf{r},0) &= 0.
     \label{eq:initial_conditions}
\end{align}
In Eq.(\ref{eq:wave_eq}), $\mathcal{D}[\cdot]$ is a differential operator expressing the PDE, and $c \in \mathbb{R}$ is the medium speed of sound, assumed to be a known constant. Since the PDE contains a second order derivative in time, two initial conditions are required. Equation (\ref{eq:initial_condition}) specifies the initial pressure field, and Eq.(\ref{eq:initial_conditions}) expresses that the initial rate of change of the pressure field is zero.

The domain is considered unbounded, with no reflected waves arriving from outside. To express this, a first-order absorptive boundary condition\cite{clayton1977absorbing} is considered
\begin{equation}
    \mathcal{B}[p] := \nabla p(\mathbf{r},t) \cdot \mathbf{n} +\frac{1}{c}\frac{\partial p}{\partial t} = 0 \quad \text{at} \ \mathbf{r}\in \partial\Omega,
    \label{eq:abs_boundary}
\end{equation}
where $\mathbf{n}$ is the unit vector normal to the boundary $\partial\Omega$. 

The goal of sound field reconstruction is to estimate the entire pressure field from noisy observations,
\begin{equation}
    \begin{aligned}
    \hat{p}_{mn} = p(\mathbf{r}_m, t_n) + e_{mn} \quad \text{for} \ m & =0,\dots,M_\text{ob}-1 \\ \text{and} \ n & =0,\dots,N-1,
    \label{eq:measurements}
    \end{aligned}
\end{equation}
where $\mathbf{r}_0, \dots, \mathbf{r}_{M_\text{ob}-1}$ are the sensor locations, $t_0, \dots, t_{N-1}$ are the time samples, and $e_{mn}$ is additive noise.
Measurements are typically performed using microphone arrays or distributed sensors. Therefore, the pressure is finely sampled over time, but only a few positions are sampled over space. 

Let $p(\mathbf{r},t; \boldsymbol{\theta})$ approximate the solution to Eqs.\ (\ref{eq:wave_eq}-\ref{eq:abs_boundary}), where $\boldsymbol{\theta}\in \mathbb{R}^{N_\theta}$ are the parameters of the approximation function. The reconstruction is formulated as solving the optimization problem
\begin{equation}
    \begin{aligned}
    &\min_{\boldsymbol{\theta}} \{\mathcal{L}_\text{data}\left(p(\mathbf{r},t_0; \boldsymbol{\theta}), \dots, p(\mathbf{r},t_{N-1}; \boldsymbol{\theta})\right) +\mathcal{L}_\text{sp}(p(\mathbf{r},t_0; \boldsymbol{\theta})) \}\\
    & \textrm{subject to} \\
    & \mathcal{D}[p(\mathbf{r},t;\boldsymbol{\theta})] = 0 \ \textrm{in} \  \Omega \times [0, T]\\
    & \mathcal{B}[p(\mathbf{r},t;\boldsymbol{\theta})] = 0 \ \textrm{in} \  \partial\Omega \times [0, T],
    \end{aligned}
    \label{eq:optimization}
\end{equation}
where $\mathcal{L}_\text{sp}$ is a sparsity constraint, and $\mathcal{L}_\text{data}$ is a measure of the difference between predictions and observations,
\begin{equation}
    \begin{aligned}
    &\mathcal{L}_\text{data}\left(p(\mathbf{r},t_0; \boldsymbol{\theta}), \dots, p(\mathbf{r},t_{n-1}; \boldsymbol{\theta})\right) = \\ 
    &\frac{1}{M_\text{ob}N}\sum_{m=0}^{M_\text{ob}-1}\sum_{n=0}^{N-1} \left| p(\mathbf{r}_m,t_n; \boldsymbol{\theta}) - \hat{p}_{mn} \right|^2.
    \end{aligned}
\end{equation}

\subsection{Sparsity constraint}
The estimation of the sound field is based only on the measured data, i.e., the initial condition $g(\mathbf{r})$ is unknown. In highly undersampled conditions, as it is often the case in sound field reconstruction, the problem cannot be solved unless additional constraints are imposed. For example, if the sound field is generated by a few sound sources, $g(\mathbf{r})$ is zero in most of the domain. To include this, a sparsity-promoting constraint  for the initial condition is considered, 
\begin{equation}
    \mathcal{L}_\text{sp}(p(\mathbf{r},t_0; \boldsymbol{\theta})) = \frac{1}{M_\text{sp}}\sum_{m=0}^{M_\text{sp}-1} \left| p(\mathbf{r}_m,t_0; \boldsymbol{\theta}) \right| = \| \mathbf{p}_\text{sp}^0 \|_1 / M_\text{sp},
    \label{eq:regu}
\end{equation}
where $\mathbf{r}_0,\dots,\mathbf{r}_{M_\text{sp}-1}$ are collocation points sampled over space, and $\mathbf{p}_\text{sp}^0 \in \mathbb{R}^{M_\text{sp}}$ represents the vector of pressure values at the these points. The constraint described in Eq.~(\ref{eq:regu}) is equivalent to the $\ell_1$-norm of the initial pressure at the collocation points. Therefore, the minimization of $\mathcal{L}_\text{sp}$ results in a sparse initial pressure in the $\ell_1$-norm sense, resulting in an optimization problem reminiscent of the lasso regression.\cite{verburg2018reconstruction} Such sparsity-promoting constraint is appropriate for estimating the initial conditions of spatially-localized acoustic sources, or problems where the pressure is mostly zero at $t=0$. 

\section{Differentiable physics}
\begin{figure}
    \centering
    \includegraphics[width=0.5\linewidth]{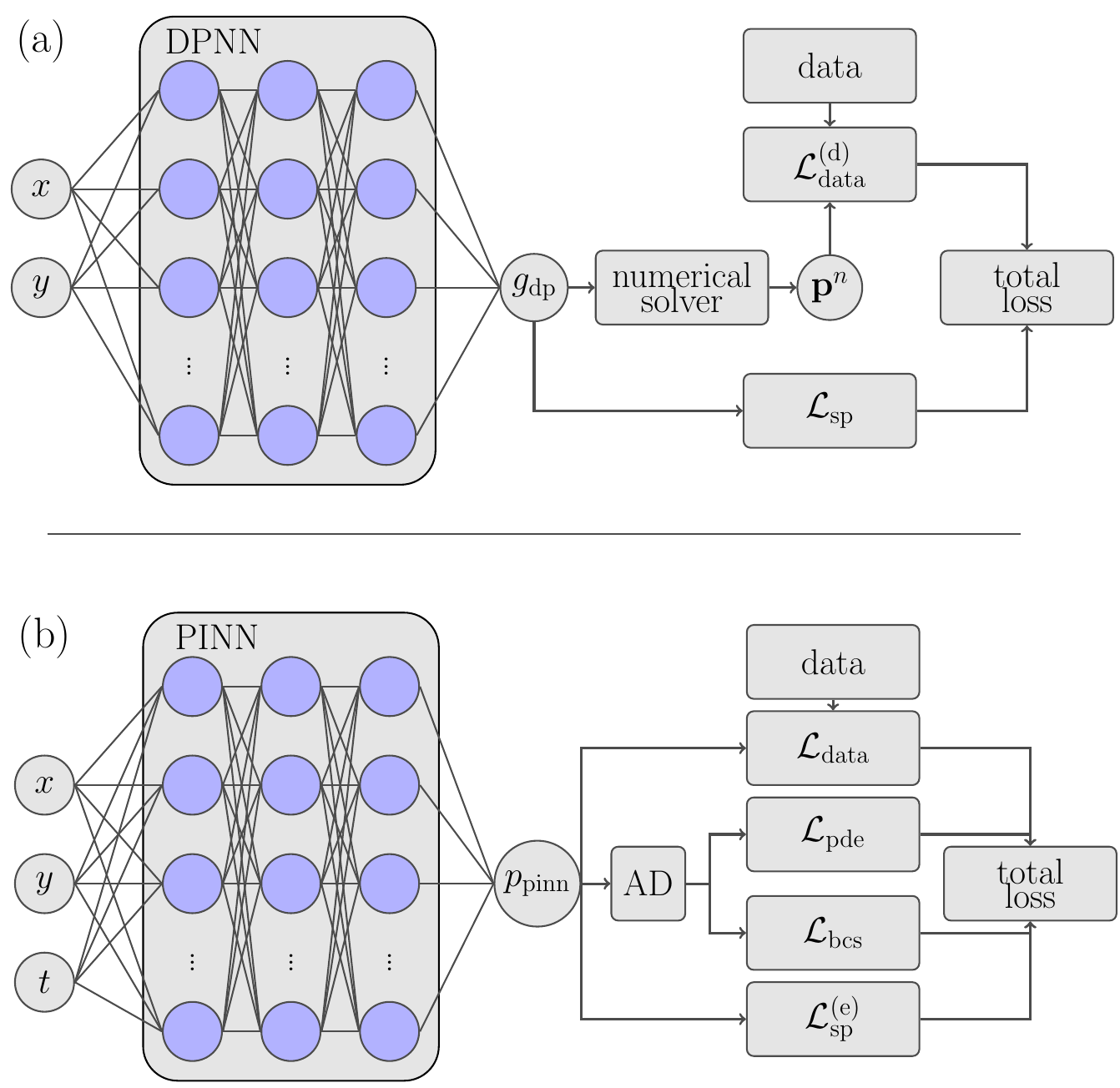}
    \caption{(a): Diagram of the DP model. The inputs to the differentiable physics neural network (DPNN) are spatial coordinates $x,y$ and the output is the estimated initial pressure $g_\text{dp}(\mbf{r};\boldsymbol{\theta})$. During training, $x,y$ are taken on a grid so that the output is $\mathbf{p}^0$, and a numerical solver is used to compute the pressure field at later times, $\mathbf{p}^1,\dots,\mathbf{p}^{n-1}$. The total loss is composed by a data fitting term $\mathcal{L}_\mathrm{data}^{\text{(d)}}$ and a regularization term $\mathcal{L}_\mathrm{reg}$. AD is used to back-propagate the loss gradient through the PDE solver and train the network. (b): Diagram of a PINN. The inputs to the network are spatio-temporal coordinates $x,y,t$ and the output is the acoustic pressure $p_\text{pinn}(\mathbf{r},t;\boldsymbol{\theta})$. AD is used for evaluating the partial derivatives in the PDE and the boundary conditions. The total loss is composed of the differential operator terms $\mathcal{L}_\mathrm{pde}$ and $\mathcal{L}_\mathrm{bcs}$, a data fitting term $\mathcal{L}_\mathrm{data},$ and a regularization term $\mathcal{L}_\mathrm{reg}^\text{(e)}$.}
    \label{fig:dp_diagram}
\end{figure}

In the proposed DP approach a neural network $g_\text{dp}(\mathbf{r};\boldsymbol{\theta})$ models the unknown initial pressure of Eq.~(\ref{eq:initial_condition}), and not the entire PDE solution (like PINNs do). The physical constraints are imposed by applying a numerical PDE solver to the network output. A diagram of the DP model is shown in Fig. \ref{fig:dp_diagram}(a). Training the DP neural network amounts to solving the optimization problem  
\begin{equation}
    \min_{\boldsymbol{\theta}} \ \left\{ \lambda_\text{data} \mathcal{L}_\text{data}^{\text{(d)}}(\mbf{p}^0,\mbf{p}^1\dots, \mbf{p}^{N-1}) + \lambda_\text{reg} \mathcal{L}_\text{reg}(g_\text{dp}(\mathbf{r};\boldsymbol{\theta})) \right\}, 
    \label{eq:dp_optimization}
\end{equation}
where $\mbf{p}^1,\dots, \mbf{p}^{N-1}$ is the numerical solution of the PDE computed in a $M_\text{grid}$-dimensional discretization grid, and $\mbf{p}^0$ is obtained by sampling the neural network $g_\text{dp}(\mbf{r};\boldsymbol{\theta})$ at the grid positions. Therefore, $\mbf{p}^n$ is a vector in $\mathbb{R}^{M_\text{grid}}$ instead of being a continuous function. 
For simplicity and without loss of generality, it is assumed that for each observation position, $\mbf{r}_0, \dots, \mbf{r}_{M_\text{ob}-1}$, there is a point in the discretization grid.
A data fitting function, $\mathcal{L}_\text{data}^{\text{(d)}}$, that operates on the discrete pressure can be expressed as
\begin{equation}
\begin{aligned}
     & \mathcal{L}_\text{data}^{\text{(d)}}\left(\mbf{p}^0, \dots, \mbf{p}^{N-1}\right) = \\
    & \frac{1}{M_\text{ob}N}\left( \left\| \mbf{M} \mbf{p}^{0}(\boldsymbol{\theta}) - \hat{\mbf{p}}^0 \right\|^2 + \sum_{n=1}^{N-1} \left\| \mbf{M} \mbf{p}^{n}(\mbf{p}^{0}(\boldsymbol{\theta})) - \hat{\mbf{p}}^n \right\|^2 \right),
\end{aligned}
\label{eq:dp_optimization_norm}
\end{equation}
where $\mbf{M}$ is a $M_\text{ob}\times M_\text{grid}$ binary matrix that extracts the pressure values at the observation positions, and $\hat{\mbf{p}}^n \in \mathbb{R}^{M_\text{ob}}$ denotes the observations in Eq. (\ref{eq:measurements}) arranged as a vector. The term for the initial pressure is taken out of the summation to emphasize that $\mbf{p}^0(\boldsymbol{\theta})$ is the network output while $\mbf{p}^{n}(\mbf{p}^{0}(\boldsymbol{\theta}))$, for $n=1,\dots,N-1$ are given by the numerical solver using $\mbf{p}^{0}(\boldsymbol{\theta})$ as initial condition.

The finite difference method \cite{langtangen2017finite} solves the PDE numerically. The solution is obtained by applying the explicit time integration scheme
\begin{subequations}\label{eq:time_integration}
\begin{align}
    &\mbf{p}^{1} = \mbf{p}^{0} + 0.5 \mbf{L} \mbf{p}^{0}, \ \text{and} \label{eq:time_integrationa}\\
    &\mbf{p}^{n+1} = 2\mbf{p}^{n} - \mbf{p}^{n-1} + \mbf{L} \mbf{p}^{n}, \ \text{for} \ n = 1,\dots,N-1,\label{eq:time_integrationb}
\end{align}
\end{subequations}
where $\mbf{L} = (c \Delta t / \Delta r)^2 \mbf{L}_\Delta$ and $\mbf{L}_\Delta \in \mathbb{R}^{M_\text{grid}\times M_\text{grid}}$ is the central difference approximation of the Laplace operator $\nabla^2[\cdot]$. The scalars $\Delta t$ and $\Delta r$ are the sampling period and grid spacing, respectively. It is assumed that the grid spacing is the same for all dimensions, e.g. in 2D, $\Delta r = \Delta x = \Delta y$. The initial step of the solver, Eq. (\ref{eq:time_integrationa}), is derived from the finite difference approximation of both initial conditions, Eqs.\ (\ref{eq:initial_condition}) and (\ref{eq:initial_conditions}).

The finite difference method provides point-wise function values that are directly compared with the observations $\hat{p}_{mn}$, and its basis functions are local, leading to sparse Jacobian matrices.\cite{xu2022physics} Nonetheless, other numerical methods, such as finite elements or spectral elements, or physical models based on other assumptions as ray tracing, can be used for DP. Ultimately, all numerical methods can be decomposed into a series of arithmetic operations that AD can differentiate.

Like any numerical solver, DP models must satisfy stability conditions to successfully solve the PDE. In the proposed finite difference-based DP model the Courant criterion $\Delta t < \Delta r / c$\cite{langtangen2017finite} must be satisfied. 
This stability requirement on the discretization grid determines the computational load of training the DP model, as each training iteration requires a forward/backward pass through every time step in Eq. (\ref{eq:time_integration}). An advantage is that the stability of numerical PDE solvers is well understood over decades of research, while for other physics-informed machine learning approaches this is still an open problem. 

Central to the proposed DP approach is the neural network, $g_\text{dp}(\mathbf{r};\boldsymbol{\theta})$, that models the initial condition, $p(\mathbf{r},0)$, as a continuous implicit function.\cite{sitzmann2020implicit} Unlike discrete models, implicit representations encode information as a continuous, smooth function that maps any input coordinate within the domain to a corresponding output value. Therefore, even if the PDE is solved on a fixed discrete grid during training, the resolution can be increased by sampling the neural network on a finer grid, and then solve the PDE with a higher resolution numerical solver. 

\subsection{Automatic differentiation}
A core aspect of DP is the use of AD to automatically compute the gradient of the loss trough the numerical solver. To train the DP model via gradient descent, the gradient of the loss function in (\ref{eq:dp_optimization}) with respect to the network parameters must be computed.
Manually deriving an expression for the gradient is possible yet cumbersome. To illustrate this, we start by applying the chain rule to Eq. (\ref{eq:dp_optimization_norm})
\begin{equation}
    \frac{\partial \mathcal{L}_\text{data}^{\text{(d)}}}{\partial \boldsymbol{\theta}} = \frac{\partial \mathcal{L}_\text{data}^{\text{(d)}}}{\partial \mbf{p}^0} \frac{\partial \mbf{p}^0}{\partial \boldsymbol{\theta}},
    \label{eq:gradient}
\end{equation}
where the Jacobian matrix of a vector valued function $\mbf{f}:\mathbb{R}^{n} \rightarrow \mathbb{R}^m$ is denoted
\begin{equation*}
    \frac{\partial \mbf{f}(\mbf{x})}{\partial \mbf{x}} = \begin{bmatrix}
    \frac{\partial f_0}{\partial x_0} & \cdots & \frac{\partial f_0}{\partial x_{n-1}} \\
    \vdots & \ddots & \vdots \\
    \frac{\partial f_{m-1}}{\partial x_0} & \cdots & \frac{\partial f_{m-1}}{\partial x_{n-1}}
\end{bmatrix}.
\end{equation*}
The second term in the product of Eq. (\ref{eq:gradient}), $\partial \mathbf{p}^0 / \partial \boldsymbol{\theta}$, is simply the derivative of the network output with respect to the network parameters. The first term can be computed by applying the chain rule recursively for each term in the sum of Eq. (\ref{eq:dp_optimization_norm}),
\begin{equation}
\begin{aligned}
    \frac{\partial \mathcal{L}_\text{data}^{\text{(d)}}}{\partial \mbf{p}^0} = & \frac{1}{M_\text{ob}N}\frac{\partial \|\mbf{Mp}^0 - \hat{\mathbf{p}}^0 \|^2}{\partial \mbf{p}^0} +\\ 
    & \frac{\partial \mathcal{L}_\text{data}^{\text{(d)}}}{\partial \mbf{p}^1}\frac{\partial \mbf{p}^1}{\partial \mbf{p}^0} + \\
    & \frac{\partial \mathcal{L}_\text{data}^{\text{(d)}}}{\partial \mbf{p}^2}\left[ \frac{\partial \mbf{p}^2}{\partial \mbf{p}^0} + \frac{\partial \mbf{p}^2}{\partial \mbf{p}^1}\frac{\partial \mbf{p}^1}{\partial \mbf{p}^0}\right] +\dots
\end{aligned}
\label{eq:gradient_chain}
\end{equation}
where the dependency of $\mbf{p}^{n}$ with respect to the previous steps is given by the update rule of the numerical solver, Eq. (\ref{eq:time_integration}). The gradient of the loss with respect to the pressure field at each time step can be obtained by differentiation of Eq. (\ref{eq:dp_optimization_norm}),
\begin{equation}
    \frac{\partial \mathcal{L}_\text{data}^{\text{(d)}}}{\partial \mbf{p}^{n}} = \frac{2}{M_\text{ob}N}(\mbf{Mp}^n - \hat{\mbf{p}}^n)^T \mbf{M} \ \text{for} \ n = 1, \dots, N-1, 
\end{equation}
and the derivatives of the pressure field at a time step with respect to the previous steps can be computed from Eq. (\ref{eq:time_integration}).

AD computes the gradient of the loss, Eq. (\ref{eq:dp_optimization_norm}), automatically and efficiently, and only the forward numerical solver is required. In a forward pass through the DP model (i.e., the neural network and solver) intermediate variables are populated and dependencies are recorded in a computational graph. Then, in a reverse pass, the gradient is calculated by propagating adjoints backward from $\mathcal{L}_\text{data}^{\text{(d)}}$ to the network parameters $\boldsymbol{\theta}$.

\subsection{Boundary conditions}
To handle the unbounded domain, the boundary condition of Eq.~(\ref{eq:abs_boundary}) is incorporated into the numerical solver. The finite difference approximation of the absorptive boundary computes the values of $\mbf{p}^{n+1}$ at the boundary based on $\mbf{p}^{n}$ at the boundary and adjacent points, as well as $\mbf{p}^{n+1}$ at the adjacent points.\cite{clayton1977absorbing} Deriving an expression for the loss gradient that includes these dependencies would be very laborious. This showcases the convenience and flexibility of using AD for differentiating though the numerical solver, since only the forward update rule derived from Eq.~(\ref{eq:abs_boundary}) is required. A first-order absorptive boundary condition is chosen for simplicity, but more precise higher-order absorptive conditions or perfectly matched layers could be implemented.

The focus of this study is solving sound field reconstruction problems in the general form, when no other information except the observed pressure data is known. For this reason, a large domain with absorptive boundary conditions is employed. If additional information about the domain geometry is available, other boundary conditions could be considered (e.g., Neumann, Robin or Dirichlet), either as part of the numerical solver or modeled and learned by the neural network.

\section{Physics-informed neural networks}\label{sec:Physics_informed_neural_networks}
PINNs are a popular and powerful physics-driven deep learning approach, which is included here as benchmark. In contrast to the proposed DP model, PINNs approximate the entire solution to Eqs.\ (\ref{eq:wave_eq}-\ref{eq:abs_boundary}) with a neural network, $p_\text{pinn}(\mathbf{r},t;\boldsymbol{\theta})$. The network inputs are position $\mathbf{r} \in \Omega$ and time $t \in [0, T]$, and the output is pressure. A diagram of a PINN is shown in Fig.\ \ref{fig:dp_diagram}(b). To train the PINN, the problem (\ref{eq:optimization}) is formulated as the unconstrained optimization problem
\begin{equation}
\begin{aligned}
    \min_{\boldsymbol{\theta}} \ \{ & \lambda_\text{data} \mathcal{L}_\text{data}(p_\text{pinn}(\mathbf{r},t_0;\boldsymbol{\theta}),\dots,p_\text{pinn}(\mathbf{r},t_{N-1};\boldsymbol{\theta})) + \\
    & \lambda_\text{sp} \mathcal{L}_\text{sp}^\text{(e)}(p_\text{pinn}(\mathbf{r},t;\boldsymbol{\theta})) +
    \lambda_\text{pde} \mathcal{L}_\text{pde}(p_\text{pinn}(\mathbf{r},t;\boldsymbol{\theta})) + \\ 
    & \lambda_\text{bcs} \mathcal{L}_\text{bcs}(p_\text{pinn}(\mathbf{r},t;\boldsymbol{\theta})) \}, 
\end{aligned}
    \label{eq:optimization_pinn}
\end{equation}
where the PDE is introduced through the soft penalty
\begin{equation}
    \mathcal{L}_\text{pde}(p_\text{pinn}(\mathbf{r},t;\boldsymbol{\theta})) = \frac{1}{M_\text{pde}}\sum_{m=0}^{M_\text{pde}-1} \left| \mathcal{D}[p_\text{pinn}(\mathbf{r}_m,t_m;\boldsymbol{\theta})] \right|^2, 
    \label{eq:loss_pde_pinn}
\end{equation}
and $(\mathbf{r}_m, t_m) \in \Omega\times[0,T]$, $m=0,\dots,M_\text{pde}-1$ are collocation points stochastically sampled at each training iteration. Samples over the spatio-temporal domain are taken to obtain $M_\text{pde}$ values of $x, y, t$ to use as network inputs in which to evaluate the PDE constraint.
Similarly, the boundary condition is introduced through the penalty
\begin{equation}
    \mathcal{L}_\text{bcs}(p_\text{pinn}(\mathbf{r},t;\boldsymbol{\theta})) = \frac{1}{M_\text{bcs}}\sum_{m=0}^{M_\text{bcs}-1} \left| \mathcal{B}[p_\text{pinn}(\mathbf{r}_m,t_m;\boldsymbol{\theta})] \right|^2, 
    \label{eq:loss_boundary_pinn}
\end{equation}
where $(\mathbf{r}_m, t_m) \in \partial\Omega\times[0,T]$, i.e, the collocation points $\mathbf{r}_m$ are sampled on the boundaries only. The operator $\mathcal{B}[\cdot]$ corresponds to first order absorptive boundary conditions expressed by Eq. (\ref{eq:abs_boundary}). Computing $\mathcal{L}_\text{pde}$ and $\mathcal{L}_\text{bcs}$ involves evaluating the partial derivatives of the network's output with respect to its inputs at the collocation points, which is done using AD. 

The function $\mathcal{L}_\text{sp}^\text{(e)}$ is the sparsity constraint term similar to Eq. (\ref{eq:regu}),
\begin{equation}
    \mathcal{L}_\text{sp}^\text{(e)}(p_\text{pinn}(\mathbf{r},t; \boldsymbol{\theta})) = \frac{1}{M_\text{sp}}\sum_{m=0}^{M_\text{sp}-1} \left| p_\text{pinn}(\mathbf{r}_m,t_m; \boldsymbol{\theta}) \right|,
    \label{eq:regu_mod}
\end{equation}
where $(\mathbf{r}_m, t_m) \in \Omega\times[0,t_\text{e}]$, and $t_\text{e}$ is an early time for which the waves have not propagated far. Imposing the sparsity constraint in the interval $[0,t_\text{e}]$ instead of just at $t=0$ helps the PINN to incorporate this prior information into the solution. Similar strategies are common to include the known initial conditions when solving the forward wave equation using PINNs.\cite{rasht2022physics} 
Note that the use of sparsity-promoting constraints in PINNs has not yet been applied to sound field reconstruction problems. Sparse constraints are nonetheless useful to achieve meaningful results when data is very scarce.\cite{chen2021physics} 

The weights $\lambda_\text{data}, \lambda_\text{sp}, \lambda_\text{pde},$ and $\lambda_\text{bcs}$ in (\ref{eq:optimization_pinn}) balance the loss terms during training. The choice of suitable weights is discussed in Section \ref{sec:lambdas}.

\section{Weighting parameters}\label{sec:lambdas}
The DP model requires to balance the loss terms $\mathcal{L}_\text{data}^\text{(d)}$ and $\mathcal{L}_\text{sp}$ through the weights $\lambda_\text{data}$ and $\lambda_\text{sp}$.
An annealing algorithm\cite{wang2021understanding} is used to select the weighting parameters. Every certain number of training iterations, the weights are updated by computing 
\begin{equation}
    \hat{\lambda}_\text{data} = \frac{\| \partial \mathcal{L}_\text{data}^\text{(d)} /\partial \boldsymbol{\theta}_n \| + \| \partial \mathcal{L}_\text{sp} /\partial \boldsymbol{\theta}_n \|}{\| \partial \mathcal{L}_\text{data}^\text{(d)} /\partial \boldsymbol{\theta}_n \|},
\end{equation}
where $\boldsymbol{\theta}_n$ are the values of the network parameters at the $n$th iteration. To avoid large update jumps a moving average of parameter $\alpha$ is used,
\begin{equation}
    \lambda_\text{data} \leftarrow \alpha\lambda_\text{data} + (1-\alpha)\hat{\lambda}_\text{data}.
    \label{eq:moving_average}
\end{equation}
A similar process is applied to update the PINN weights $\lambda_\text{data}, \lambda_\text{sp}, \lambda_\text{pde}$ and $\lambda_\text{bcs}$. Such an algorithm was proposed to balance the back-propagated gradients during the training of PINNs.\cite{wang2021understanding}

\section{Numerical experiments}
Numerical experiments are performed that compare the proposed DP model with PINNs for sound field reconstruction problems. A 2+1D domain is defined, where the spatial domain is a square of side length $L=1$, the temporal domain has a duration of $T=0.343$, and the speed of sound is $c=1$. These are the dimensionless physical variables that result from scaling a domain with $L=1$ m, $T=1$ ms, and $c=343$ m/s. The temporal domain is divided into $n=50$ samples, giving a sampling period $\Delta t = 7.0\times 10^{-3}$. 

For the DP finite difference solver a regular discretization grid of $M_\text{grid}=100^2$ is defined, therefore $\Delta r\approx 1.0\times 10^{-2}$. This discretization satisfies the stability criterion $\Delta t < \Delta r/ (\sqrt{2}c)$.\cite{langtangen2017finite}

For the PINN, $L=2\times10^3$ collocation points in $ \Omega \times [0,T]$ are used for fitting the PDE [Eq. (\ref{eq:loss_pde_pinn})], $P=8\times10^2$ collocation points in $ \partial\Omega \times [0,T]$ are used for fitting the boundary condition [Eq. (\ref{eq:loss_boundary_pinn})], and $M=2\times10^2$ points in $\Omega \times [0,0.1T]$ are used to fit the sparsity-promoting constraint [Eq. (\ref{eq:regu_mod})]. The collocation points are drawn from Sobol quasi-random sequences at each training step.

The DP and PINN architectures are shown in Table \ref{tab:architecture}. 
The ADAM optimizer with a learning rate of $10^{-4}$ is used for training both networks. The number of optimizer steps, $5\times 10^4$ for the DP network and $5\times 10^5$ for the PINN, is chosen heuristically by observing the convergence of the total loss [see Fig. \ref{fig:pulse_training}(e)]. Each of the networks is trained in 1-3 hours on an NVIDIA RTX A2000 GPU. The number of training iterations per second on average was 10 for DP and 35 for PINN. The models are implemented in PyTorch\cite{paszke2019} and available at the repository: \url{https://github.com/samuel-verburg/differentiable-soundfield-reconstruction.git}

\begin{table}[]
\begin{tabular}{lcccc}
\hline
Neural network                            & Activation & \# layers & \# units per layer \\ \hline
PINN $p_\text{pinn}(\mathbf{r},t;\theta)$ & periodic (SIREN)\cite{sitzmann2020implicit}      & 4         & 128 \\
DPNN $g_\text{dp}(\mathbf{r};\theta)$     & periodic (SIREN)      & 3         & 64 \\ \hline
\end{tabular}
\caption{Neural networks architecture.}
\label{tab:architecture}
\end{table}

\begin{figure}
    \centering
    \includegraphics[width=0.5\linewidth]{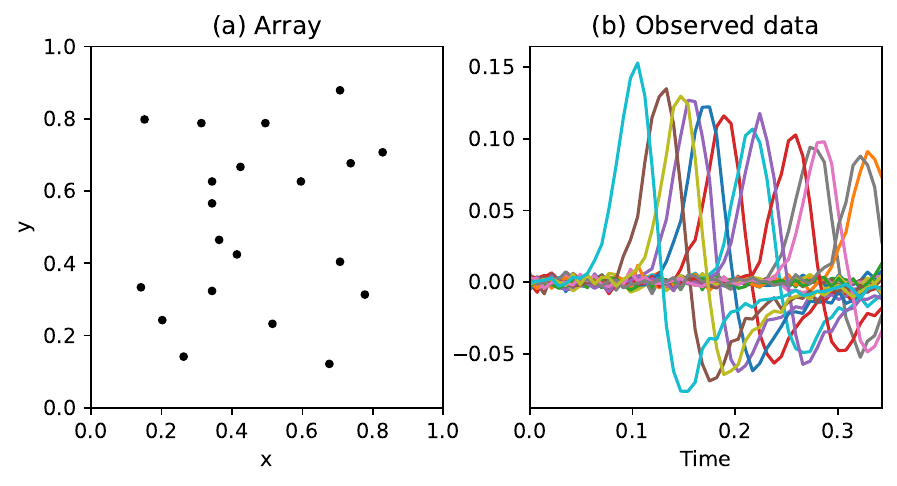}
    \caption{Left: Measurement positions used for estimating the sound field. Right: simulated noisy data at each of the measurement positions over time for a sound field consisting of a single pulse at the center of the domain.}
    \label{fig:pulse_data}
\end{figure}

\begin{figure*}
    \centering
    \includegraphics[width=0.7\linewidth]{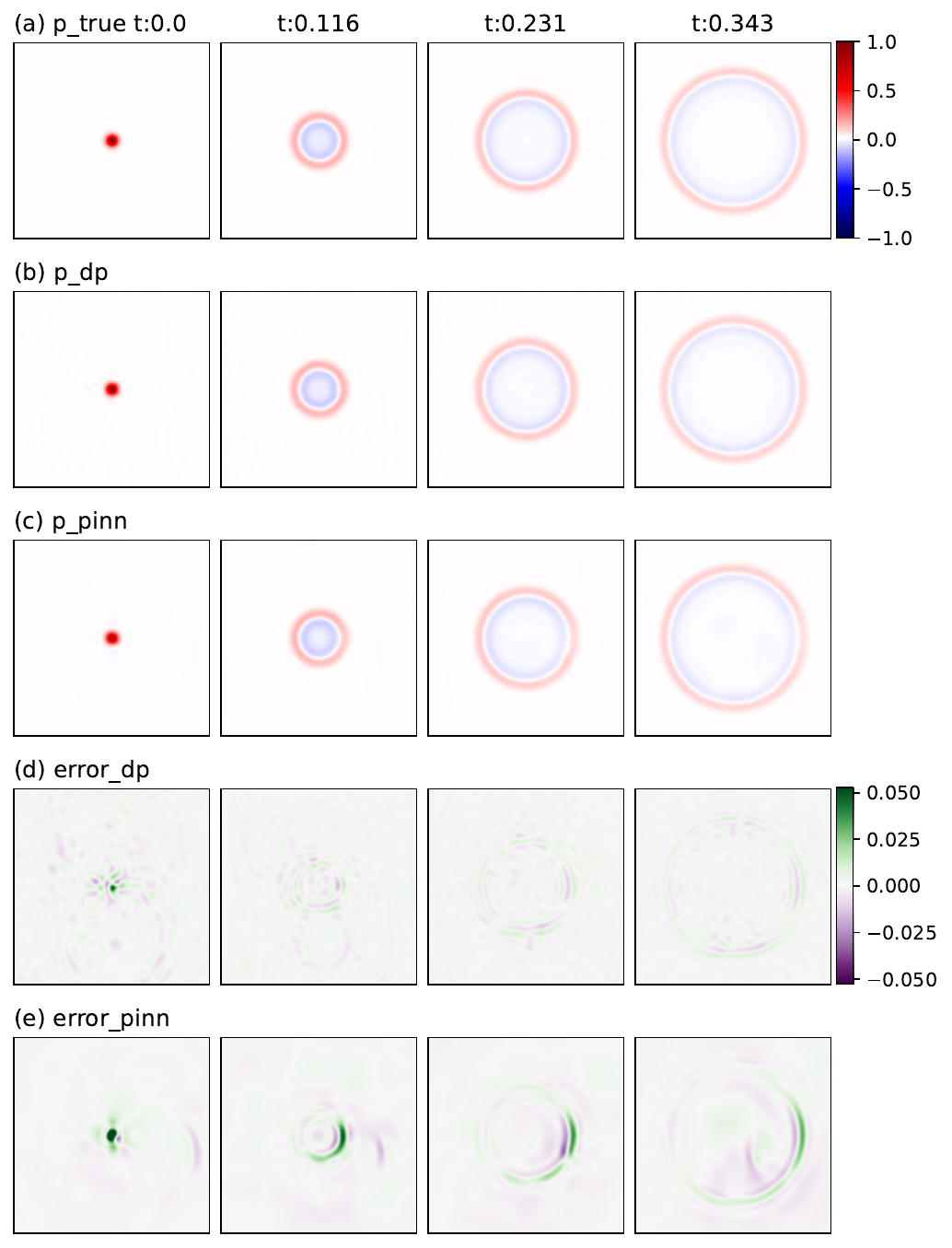}
    \caption{Sound field consisting on a single pulse at the domain center. Each column corresponds to a time frame. Row (a): reference solution. Row (b): DP model estimation. Row (c): PINN estimation.
    Row (d): difference between reference and DP. Row (e): difference between reference and PINN.}
    \label{fig:pulse_comparison}
\end{figure*}

\subsection{Single pulse}\label{sec:single_pulse}
Synthetic data for multiple sound fields is generated. For the first one, the initial condition is a single Gaussian pulse of unit amplitude and scale $\sigma=0.02$ placed at the center of the domain,
\begin{equation}
    g(\mathbf{r}) = \text{exp}\left( -0.5 \|\mathbf{r}-\mathbf{r}_0\|^2/\sigma^2 \right),
\end{equation}
where $\mathbf{r}_0 = (L/2, L/2)$. The observations used for the reconstruction conform a pseudo-random array, shown in Fig. \ref{fig:pulse_data}(a), where the sensor locations are sampled from the discretization grid within $[0.1L, 0.9L]\times[0.1L, 0.9L]$ and with a minimum distance of $0.05$ between sensors. The number of time samples is $n=50$ and the number of sensors is $m=20$. Figure \ref{fig:pulse_data}(b) shows the observed data over time. Additive Gaussian noise is added to the data such that the SNR is 20 dB.

The observed data and reference values of the sound field are obtained from the analytical solution of the acoustic wave equation in free field with a Gaussian pulse as initial condition (see Appendix \ref{sec:appendix}).
The reference sound field is computed on a grid of twice the spatial and temporal resolutions of the DP finite difference grid, i.e., $\Delta r_\text{reference}\approx 5.0\times 10^{-3}$ and $\Delta t_\text{reference}\approx 3.5\times 10^{-3}$. The mean squared error normalized by the total energy of the reference field,
\begin{equation}
    \text{NMSE} = \frac{\sum_{i} | p_\text{model}(\mathbf{r}_i, t_i) - p_\text{true}(\mathbf{r}_i, t_i) |^2}{\sum_{i} | p_\text{true}(\mathbf{r}_i, t_j) |^2},
\end{equation}
is used as error metric, where $(\mathbf{r}_i, t_i)$ are the all the spatio-temporal points on the evaluation grid, $p_\text{true}$ is the reference field, and $p_\text{model}$ is the evaluated model (either DP or PINN).

Figure \ref{fig:pulse_comparison} shows the reference sound field, the estimations of DP and PINN models, and the difference between the reference and estimations for the single Gaussian pulse. The DP results [rows (b) and (d) in Fig. \ref{fig:pulse_comparison}] show an accurate reconstruction throughout the spatio-temporal domain, with only noticeable differences at $t=0$. The NMSE is $5.3 \times 10^{-3}$. The PINN results [rows (c)  and (e) in Fig. \ref{fig:pulse_comparison}] show errors around the right side of the wavefront at all times.
It is likely that the PINN reached a local minium during training, and could not minimize the error further. The mean squared error for the PINN is $2.1\times 10^{-2}$.

\begin{figure}
    \centering
    \includegraphics[width=0.5\linewidth]{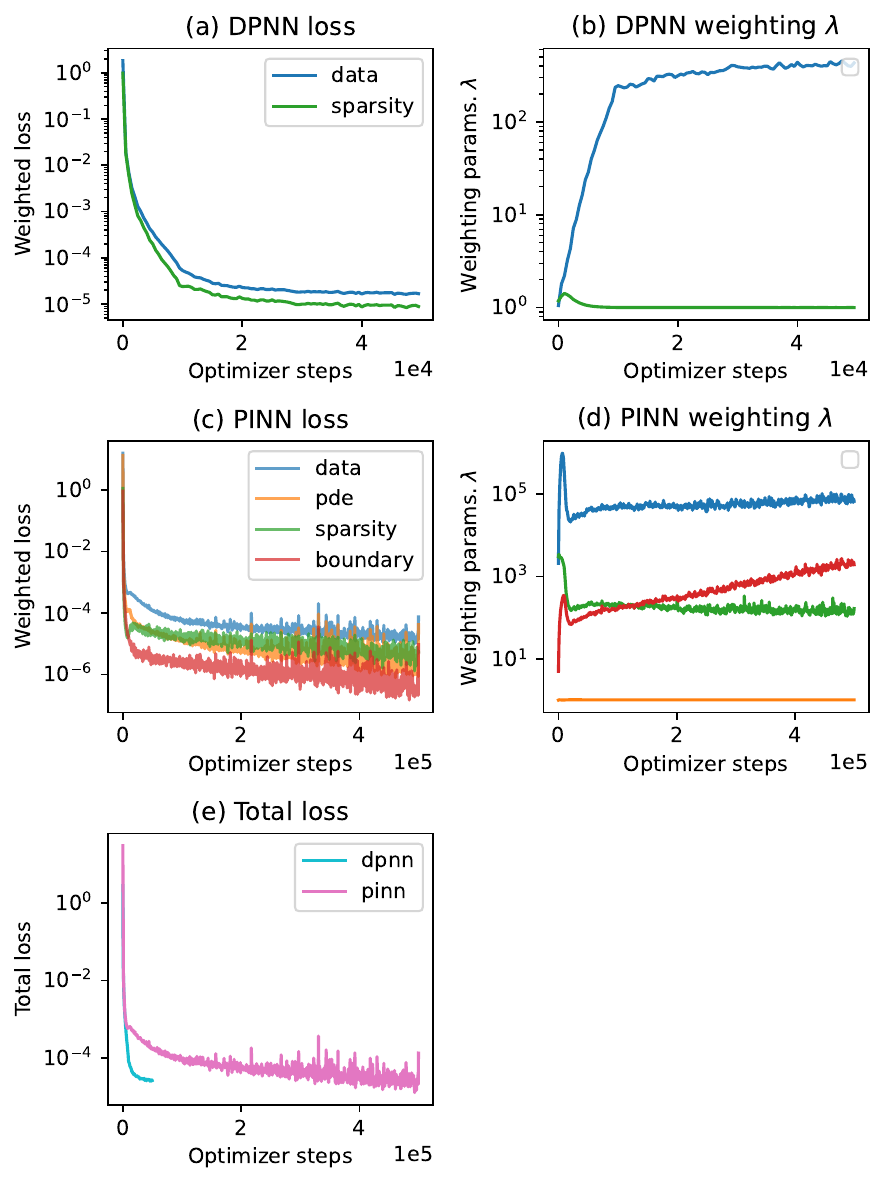}
    \caption{Dynamics of the DP neural network and PINN trained to learn the single pulse sound field. (a): Weighted loss terms, `data' stands for $\lambda_\text{data} \mathcal{L}_\text{data}^{\text{(d)}}$, and `sparsity' for $\lambda_\text{reg} \mathcal{L}_\text{reg}$. (b): Weighting parameters $\lambda_\text{data}$ and $\lambda_\text{reg}$. (c): Weighted loss terms, `data' stands for $\lambda_\text{data} \mathcal{L}_\text{data}$, `pde' for $\lambda_\text{pde} \mathcal{L}_\text{pde}$, `sparsity' for $\lambda_\text{reg} \mathcal{L}_\text{reg}$, and `boundary' for $\lambda_\text{bcs} \mathcal{L}_\text{bcs}$. (d): Weighting parameters $\lambda_\text{data}, \lambda_\text{pde}$, $\lambda_\text{reg}$, and $\lambda_\text{bcs}$. (e): Total weighted loss.}
    \label{fig:pulse_training}
\end{figure}

The training dynamics of the DP neural network and the PINN are shown in Fig. \ref{fig:pulse_training}. 
Figure \ref{fig:pulse_training}(a) shows the evolution of the loss terms $\lambda_\text{data} \mathcal{L}_\text{data}^{\text{(d)}}$ and $\lambda_\text{reg} \mathcal{L}_\text{reg}$ for the DP neural network, while Fig. \ref{fig:pulse_training}(c) shows the loss terms $\lambda_\text{data} \mathcal{L}_\text{data}$, $\lambda_\text{pde} \mathcal{L}_\text{pde}$, $\lambda_\text{reg} \mathcal{L}_\text{reg}$ and $\lambda_\text{bcs} \mathcal{L}_\text{bcs}$ for the PINN.
Since the DP model does not include a PDE term and the differential operator is computed with a stable numerical method, the training of the DP neural network is more stable. Conversely, the PINN loss function consists of many competing terms, making training much harder. 
This difference can also be seen in the weighting parameters $\lambda$ obtained from the annealing algorithm of Section \ref{sec:lambdas}, shown in Figures \ref{fig:pulse_training}(b) and (d). For the PINN, the PDE term is always assigned a weight of one, while the weights for the data, sparsity, and boundary terms range from $1\times 10^2$ to $1\times 10^5$ approximately. This indicates that the network parameters are very sensitive to the differential operator in the PDE residual, which might cause overall slow convergence and the failure to learn the appropriate solution. Finally, Figure \ref{fig:pulse_training}(e) compares the total loss, the sum of loss terms, showing that the DP neural network achieves convergence in a fraction of the optimization steps. 

\begin{figure*}
    \centering
    \includegraphics[width=1\linewidth]{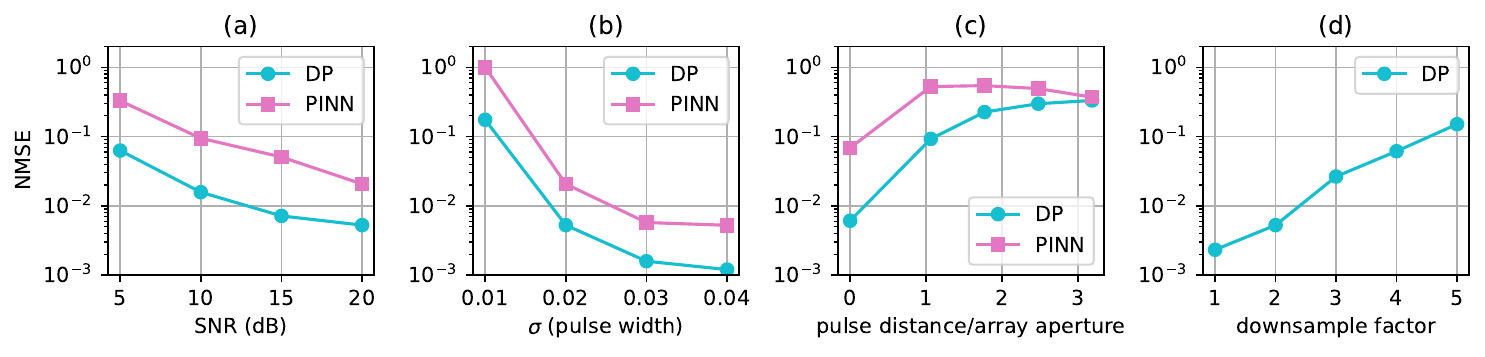}
    \caption{Normalized Mean squared error as a function of (a) the SNR, (b) the source width, (c) the pulse distance to the array center, and (d) the downsample factor between the evaluation and training grids.}
    \label{fig:nmse_vs_snr}
\end{figure*}

The reconstruction performance is analyzed by training the models in different scenarios. Figure \ref{fig:nmse_vs_snr}(a) shows the NMSE for various noise levels. The proposed DP model largely outperforms the PINN for all tested SNRs, presenting errors almost one order of magnitude smaller. 

Figure \ref{fig:nmse_vs_snr}(b) shows the NMSE as a function of the pulse scale $\sigma$, which is directly related to the frequency content of the acoustic field. The PINN fails to reconstruct the sound field of highest frequency (smallest $\sigma$), presenting a NMSE close to 1. PINNs often fail to solve PDE containing high frequencies due to spectral bias.\cite{wang2021eigenvector} 
The DP model consistently achieves lower NMSE. Since the DP network approximates only the initial pressure (and not the entire PDE solution), its representational power can focus on this simpler function, somewhat circumventing the spectral bias. It is worth noting that the DP network has one layer less and half the number of units per layer that the PINN (see Table \ref{tab:architecture}).

Figure \ref{fig:nmse_vs_snr}(c) shows the NMSE as a function of the distance between the pulse and the array center normalized by the array aperture. The experiment serves to evaluate the extrapolation capabilities of the models to areas where there is no observed data and the estimation relies only on the physics of wave propagation. The results show a degraded performance when the pulse is further away because the wavefront seen by the array is increasingly plane, making it difficult to capture its curvature and the source range. The PINN model presents a large error as soon as the source is outside the array aperture as the physics are only included as a weak constraint. Conversely, the DP model output stratifies the underlying physics by design. The error of the DP model increases more progressively, indicating a better extrapolation capabilities.  

Figure \ref{fig:nmse_vs_snr}(d) shows the NMSE for different discretization grids used during training the DP model. The x-axis corresponds to the downsample factor of the training grid with respect to the reference grid (e.g. a downsample factor of 5 means that the grid resolution for training is 5 times coarser that the one used for evaluation). Since the initial conditions are approximated with a continuous function (the DP network) we can upscale the estimation to any desired resolution. As expected, the NMSE increases for lower training resolutions, which is caused by the accumulation of numerical errors and the fact that coarser grids are not able to represent high spatial frequencies present in the initial condition. Therefore, care must be taken to select an appropriate grid during training. 
Nonetheless it is possible to obtain estimations with errors below $10^{-1}$ even when the network is trained with coarse grids. 

\begin{figure*}
    \centering
    \includegraphics[width=0.7\linewidth]{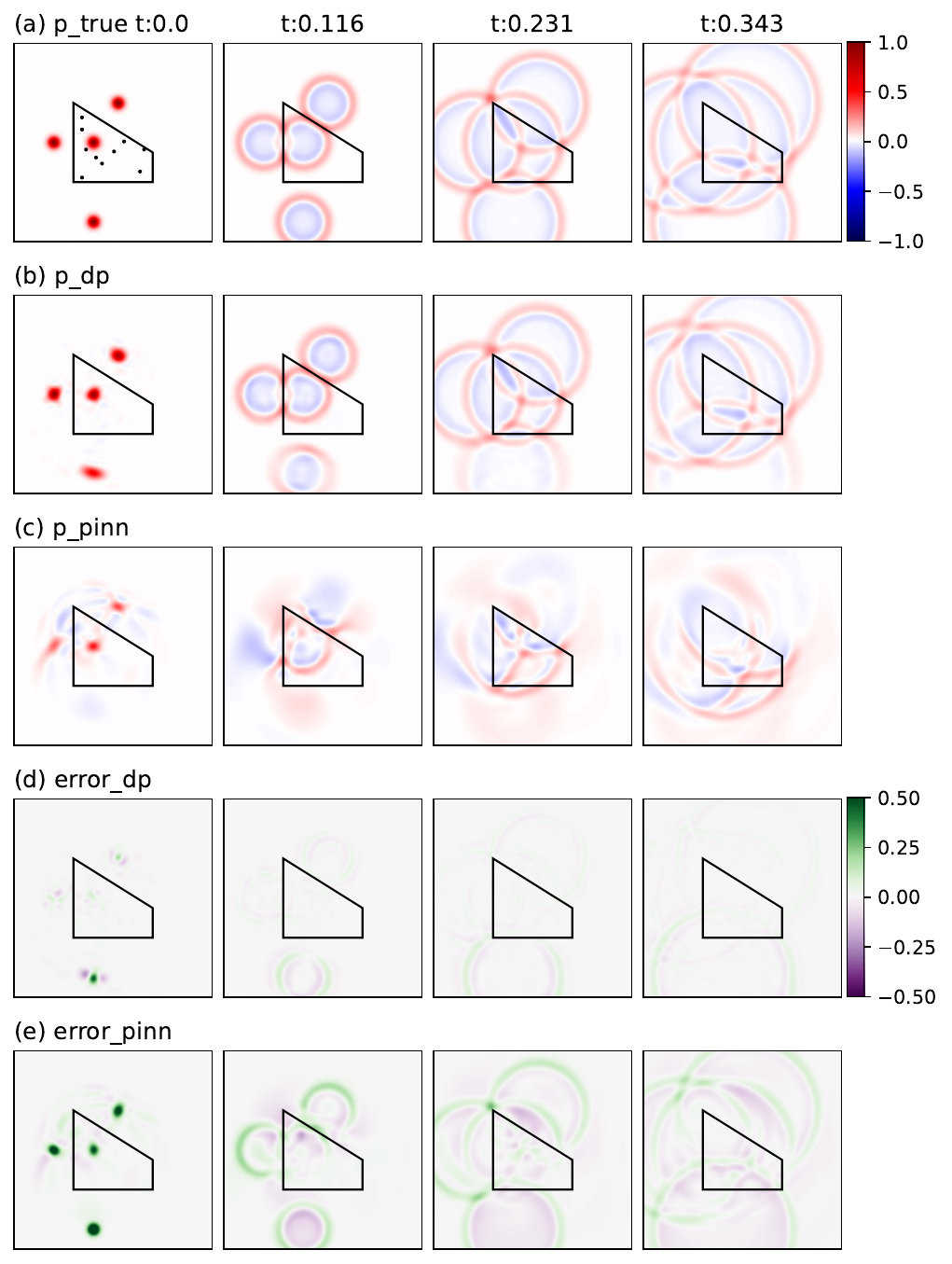}
    \caption{Sound field inside a trapezoidal enclosure with rigid walls. Each column corresponds to a time frame. Row (a): reference solution. Row (b): DP model estimation. Row (c): PINN estimation.
    Row (d): difference between reference and DP. Row (e): difference between reference and PINN.}
    \label{fig:reflective_comparison}
\end{figure*}

\subsection{Reverberant field}
In the second experiment, a reverberant field inside an enclosure is simulated. This is a typical scenario for sound field reconstruction problems in rooms and enclosures, where the goal is to estimate the pressure inside a room from a small number of distributed sensors, without knowledge of the source position, room geometry, or surface acoustic impedance. The reverberant field is simulated by modeling the direct sound and three first-order reflections as Gaussian pulses with unit amplitude and $\sigma=0.02$. Their position correspond to the image sources in a trapezoidal room with corners at $(0.3,0.3), (0.3,0.7), (0.7,0.45), (0.7,0.3)$. In this experiment the sensors locations are restricted to the interior of the trapezoid to simulate room impulse responses measured with a distribution of sensors. The number of sensors is reduced to $m=10$, and their location is indicated at the top-left panel of Fig. \ref{fig:reflective_comparison}. 

The training process is similar to the previous experiments. The DP and PINN models estimate the pressure in the same unbounded domain (i.e., the square domain with absorptive boundaries) from the observed data only. We emphasize that the models have no knowledge of the geometry and surface impedance of the enclosure.

The reference and estimation results are in Fig. \ref{fig:reflective_comparison}. The proposed DP model can accurately recover the reverberant field. Remarkably, the DP model correctly extrapolates the pressure to areas without observed data. Errors in the extrapolation are present for the furthest image source (at the bottom of the domain). That is because the wavefront of this source has not yet reached all sensors at $t=T$, therefore, the source has to be estimated from even fewer data. Nevertheless, a close agreement with the reference field is achieved across the entire domain. 

In contrast, the PINN fails to recover the pressure field, both within and outside the enclosure. This reconstruction problem is very challenging because, while the sensors only cover the enclosure's area, the field has to be estimated in the entire domain. Therefore, the extrapolation relies only on the physics of wave propagation, which PINNs incorporate as soft penalties. This seems to explain why the PINN does not estimate the pressure field outside the enclosure. In particular, it can be seen that the reflection from the bottom surface is completely missing from the PINN estimation.
In addition, the pulses at $t=0$ are smeared and their amplitude is underestimated. High frequencies, such as those in the initial pulses, are challenging for neural networks.\cite{wang2021eigenvector}
Even though the chosen SIREN architecture should help overcome the spectral bias, the PINN still struggles to learn the initial pressure.

\subsection{Complex source distributions}
\begin{figure}
    \centering
    \includegraphics[width=0.5\linewidth]{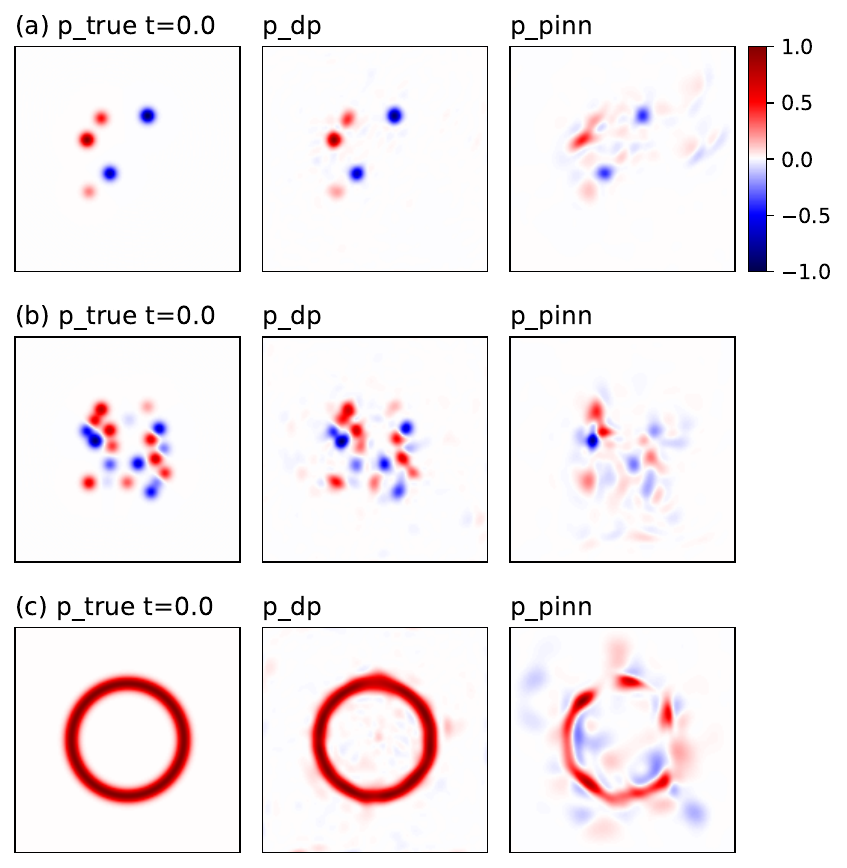}
    \caption{Initial conditions for different sound fields. The columns correspond to the reference, DP estimation, and PINN estimation, respectively. Row (a) five pulses. Row (b) twenty pulses. Row (c) continuous ring-like source.}
    \label{fig:five_pulses}
\end{figure}

Additional sound fields are synthesized to test the methods in other challenging sound field reconstruction problems, in particular when sparsity cannot be assumed. Figure \ref{fig:five_pulses}(a) and (b) correspond to the combination of five and twenty Gaussian pulses, respectively. The pulses are randomly distributed in $[0.3L, 0.7L]$ with random amplitudes in $[-1, 1]$ and $\sigma=0.02$. The number and location of the sensors and all other parameters are kept identical to the experiment with a single pulse (Sec. \ref{sec:single_pulse}). The reference is shown in the first column of Fig. \ref{fig:five_pulses}, while the DP and PINN estimations are shown in the second and third columns, respectively. The PINN clearly fails to learn a meaningful solution, while the DP is able to recover the initial pressure in both cases. The DP model's ability to recover such a complex sound field with as few as 20 sensors is remarkable. Assuming that a characteristic wavelength of the sound field is $2\sigma$ and considering that $L/\sigma=50$, a grid of $50^2$ sensors would be required to recover the sound field according to classical sampling theory (without sparsity constraint).

Figure \ref{fig:five_pulses}(c) shows a ring-like initial condition given by 
\begin{equation}
g(\mathbf{r}) = \text{exp}\left( -0.5 \left(\|\mathbf{r}-\mathbf{r}_0\|-R\right)^2/\sigma^2 \right)
\end{equation}
where $R=0.25$. In this case, the observed and reference data are obtained from a high resolution finite difference simulation. The DP model can reconstruct the sound field even when the source is a continuous line, and the initial pressure is not as sparse. 

\begin{figure}
    \centering
    \includegraphics[width=0.5\linewidth]{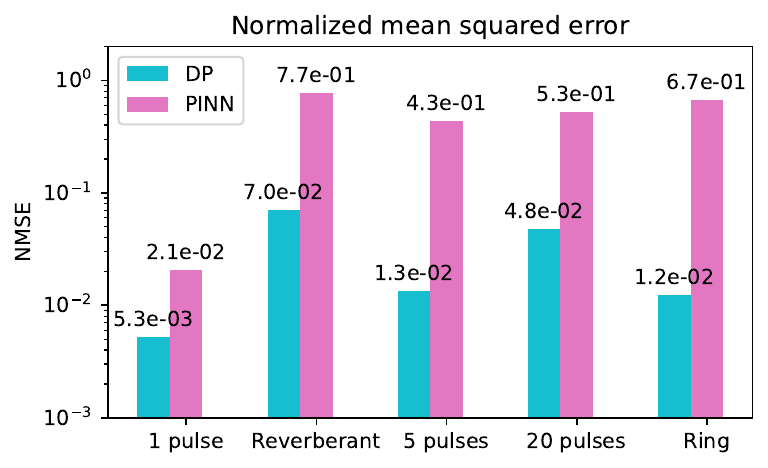}
    \caption{Normalized mean squared error for the sound fields considered in the numerical experiment.}
    \label{fig:mean_squared_error}
\end{figure}

The mean squared error of the DP and PINN models for the sound fields considered in the numerical experiments is shown in Figure \ref{fig:mean_squared_error}. The error is computed with respect to the reference field in the entire domain at the high-resolution grid (i.e., at twice the spatial and temporal resolution used for training the DP neural network). The proposed DP model presents an error approximately one order of magnitude smaller than the PINN for all cases.

\section{Conclusion}
We propose a differentiable physics approach for sound field reconstruction. Integration of a numerical solver in the training of a neural network enables the incorporation of hard physical constraints robustly. The optimization is more stable than in conventional PINNs, and convergence is achieved in a fraction of the optimizer steps. Formulating the solver in a differentiable way using AD makes the training process very simple since only the forward solver is required. 

The flexibility offered by the neural network enables us to represent complex initial conditions. Effectively, the network is an implicit neural representation of the initial pressure, in which every input coordinate is mapped to an output value. Therefore, the DP approach is generalizable beyond the training discretization, and the solutions obtained can be scaled to higher resolutions. Additionally, incorporating a sparsity-promoting constraint enables the reconstruction of sound fields with very little data. The experiments show that the DP model achieves accurate reconstructions and low errors even in challenging, highly undersampled problems.

While this study focuses on estimating a pressure field from sparse observations when no other information is available, DP is not limited to this application, and other estimation problems in acoustics can benefit from it. Absorptive boundary conditions are considered, yet a promising direction is the integration of the problem's geometry and boundary impedance, either as additional information to help reconstructing the sound field, or as additional unknowns to be estimated. Levering the AD capabilities, other parameters of the PDE could be estimated in the same DP framework. In sound field reconstruction applications it is normally assumed that the wave speed $c$ is constant and known, yet it could be included as a learnable parameter, or modeled with an additional neural network if the medium is heterogeneous. Other relevant problems in sound field reconstruction, such as optimal sensor placement,\cite{koyama2020optimizing,verburg2024optimal} could be addressed using the proposed DP framework. In conclusion, the proposed DP approach presents a promising direction for advancing spatio-temporal processing, sound field reconstruction and estimation problems in acoustics.

\acknowledgments
This work was supported by a research grant from the
DFF foundation (Grant No. 10.46540/3105-00199B).
\section*{Author declarations}
The authors declare no conflict of interest.
\section*{Data availability}
The data that support the findings of this study are available from the corresponding author upon request.

\appendix
\section{Gaussian Pulse}\label{sec:appendix}
Two-dimensional sound fields where the initial pressure distribution is a Gaussian pulse of scale $\sigma^2$ centered at $\mathbf{r}_0=(x_0,y_0)$ and with amplitude $A$, i.e.,
\begin{equation}
    g(\mathbf{r}) = A\text{exp}\left( -0.5 \|\mathbf{r}-\mathbf{r}_0\|^2/\sigma^2 \right),
\end{equation}
are considered. The analytical solution in a homogeneous quiescent medium of normalized sound speed $c=1$ is\cite{tam1993dispersion} 
\begin{equation}
    p(x,y,t) = A\sigma^2 \int_0^\infty \exp\left[{-\frac{(\xi\sigma)^2}{2}}\right]\cos(\xi t)J_0(\xi r) \xi d\xi,
    \label{eq:Gaussian_field}
\end{equation}
where $r=\|\mathbf{r}-\mathbf{r}_0\|$ and $J_0$ is the Bessel function of order zero. The integral in Eq. (\ref{eq:Gaussian_field}) is numerically evaluated using SciPy's\cite{virtanen2020scipy} integration functionality based on QUADPACK.\cite{piessens2012quadpack} 

\bibliography{main.bib}





\end{document}